\documentclass[10pt,twocolumn,secnumarabic,amssymb, nobibnotes, aps, prd]{revtex4-2}

\usepackage{xcolor}

\usepackage{graphicx}
\usepackage[utf8]{inputenc}
\usepackage{lipsum}
\usepackage{subfigure}
\usepackage{amsmath}
\usepackage{wrapfig}
\usepackage{xparse,xcoffins}
\DeclareUnicodeCharacter{0301}{\"o}
\ExplSyntaxOn
\NewCoffin\imagecoffin
\NewCoffin\labelcoffin

\keys_define:nn { miguel/label }
 {
  label   .tl_set:N = \l_miguel_label_tl,
  labelbox .bool_set:N = \l_miguel_label_box_bool,
  labelbox .default:n = true,
  fontsize .tl_set:N = \l_miguel_label_size_tl,
  fontsize .initial:n = \footnotesize,
  pos .choice:,
  pos/nw .code:n = \tl_set:Nn \l_miguel_label_pos_tl { left,up },
  pos/ne .code:n = \tl_set:Nn \l_miguel_label_pos_tl { right,up },
  pos/sw .code:n = \tl_set:Nn \l_miguel_label_pos_tl { left,down },
  pos/se .code:n = \tl_set:Nn \l_miguel_label_pos_tl { right,down },
  pos/n .code:n = \tl_set:Nn \l_miguel_label_pos_tl { hc,up },
  pos/w .code:n = \tl_set:Nn \l_miguel_label_pos_tl { left,vc },
  pos/s .code:n = \tl_set:Nn \l_miguel_label_pos_tl { hc,down },
  pos/e .code:n = \tl_set:Nn \l_miguel_label_pos_tl { right,vc },
  pos .initial:n = nw,
  unknown .code:n   = \clist_put_right:Nx \l_miguel_label_clist
                       { \l_keys_key_tl = \exp_not:n { #1 } }
 }
\clist_new:N \l_miguel_label_clist
\box_new:N \l_miguel_label_box
\box_new:N \l_miguel_label_image_box

\NewDocumentCommand{\xincludegraphics}{O{}m}
 {
  \group_begin:
  \tl_clear:N \l_miguel_label_tl
  \clist_clear:N \l_miguel_label_clist
  \keys_set:nn { miguel/label } { #1 }
  \tl_if_empty:NTF \l_miguel_label_tl
   {
    \miguel_includegraphics:Vn \l_miguel_label_clist { #2 }
   }
   {
    \SetHorizontalCoffin\imagecoffin
     {
      \miguel_includegraphics:Vn \l_miguel_label_clist { #2 }
     }
    \SetHorizontalCoffin\labelcoffin
     {
      \raisebox{\depth}
       {
        \bool_if:NTF \l_miguel_label_box_bool
         { \fcolorbox{white}{white}{\l_miguel_label_size_tl\l_miguel_label_tl} }
         { \l_miguel_label_size_tl\l_miguel_label_tl }
       }
     }
    \SetVerticalPole\imagecoffin{left}{3pt+\CoffinWidth\labelcoffin/2}
    \SetVerticalPole\imagecoffin{right}{\Width-3pt-\CoffinWidth\labelcoffin/2}
    \SetHorizontalPole\imagecoffin{up}{\Height-3pt-\CoffinHeight\labelcoffin/2}
    \SetHorizontalPole\imagecoffin{down}{3pt+\CoffinHeight\labelcoffin/2}
    \use:x{\JoinCoffins\imagecoffin[\l_miguel_label_pos_tl]\labelcoffin[vc,hc]}
    \TypesetCoffin\imagecoffin
   }
   \group_end:
 }
\NewDocumentCommand{\setlabel}{m}
 {
  \keys_set:nn { miguel/label } { #1 }
 }

\cs_new_protected:Nn \miguel_includegraphics:nn
 {
  \includegraphics[#1]{#2}
 }
\cs_generate_variant:Nn \miguel_includegraphics:nn { V }

\ExplSyntaxOff

\makeatletter
\newcommand{\twocolumncaption}{\@dblarg\@twocolumncaption}
\def\@twocolumncaption[#1]#2{%
  \renewcommand{\@makecaption}[2]{%
    \par\vskip\abovecaptionskip\begingroup\tiny\rmfamily
    \splittopskip=0pt
    \setbox\@tempboxa=\vbox{
      \@arrayparboxrestore \let \\\@normalcr
      \hsize=.5\hsize \advance\hsize-1em
      \let\\\heading@cr
      \noindent ##1\ ##2\par
    }%
    \vbadness=10000
    \setbox\z@=\vsplit\@tempboxa to .55\ht\@tempboxa
    \setbox\z@=\vtop{\hrule height 0pt \unvbox\z@}
    \setbox\tw@=\vtop{\hrule height 0pt \unvbox\@tempboxa}
    \noindent\box\z@\hfill\box\tw@\par
    \endgroup\vskip \belowcaptionskip
  }%
  \setlength{\abovecaptionskip}{4ex}%
  \caption[#1]{#2}%
}
\begin{document}
\title{Synergistic Effects of Spin-Orbit Coupling and Intercomponent Interactions in Two-Component (2+1)D Photonic Fields}

\author{Suri Deekshita}%
\author{S. Sanjay}
\author{S. Saravana Veni}%
\email {s\textunderscore saravanaveni@cb.amrita.edu.in}
\affiliation{Department of Physics, Amrita School of Physical Sciences, Amrita Vishwa Vidyapeetham, Coimbatore, 641112, Tamil Nadu, India}
\author{Conrad B. Tabi}%
\affiliation{Department of Physics and Astronomy, Botswana International University of Science and Technology, Private Mail Bag 16 Palapye, Botswana}
\author{Timol\'eon C. Kofané}
\affiliation{Department of Physics and Astronomy, Botswana International University of Science and Technology, Private Mail Bag 16 Palapye, Botswana}

\begin{abstract}
The study investigates the formation, stability and dynamic advancement of two-dimensional vortex quantum droplets within binary Bose-Einstein condensates (BECs), shaped by the interplay of photonic spin-orbit coupling (SOC) and quantum fluctuation effects. SOC leads to significant droplet stretching, resulting in vortex clusters forming in each component. The competition between photonic SOC and Lee-Huang-Yang (LHY) interactions introduces vortices into the condensate, described by the numerically solved Gross-Pitaevskii equation (GPE). The results show that droplets like structures arise at low SOC strengths and interaction parameters. The transition to vortex takes place as the SOC increases. Enhanced interactions give rise to the emergence of quantum droplets as the vortices dissipate, demonstrating fascinating dynamics. These findings enhance understanding of the physical properties of photonic SOC coupled binary BECs in 2D with LHY correction, impacting cold-atom physics and condensed matter research. The study can also be expanded to explore quantum droplets with a small atom count, which is advantageous for experimental applications.
\end{abstract}

\flushbottom
\maketitle
%
%
\thispagestyle{empty}


\section*{Introduction}
Ultracold atoms provide a fertile platform to investigate active research frontiers such as supersolidity 
\cite{recati2023supersolidity,bland2022two}, quasi-particles 
\cite{massignan2014polarons}, and quantum droplets \cite{luo2021new,tabi2023modulational}.  
Bose-Einstein condensates (BECs) are a macroscopic quantum state of matter where the quantum droplets (QDs) may be stable not only in their fundamental state but also in excited forms, such as vortices or multipole-mode quantum droplets in both two-dimensional (2D) \cite{PhysRevA.98.063602,article,PhysRevA.106.053303} and three-dimensional (3D) \cite{PhysRevA.98.013612,DONG2023113728} geometries. BECs represent one of the most intriguing states of matter, providing a macroscopic quantum system that can be meticulously controlled and studied in laboratory settings~\cite{PhysRevA.106.033309}. QDs are self-bound states that arise from a delicate balance between attractive and repulsive forces within the system, as observed in dipolar gases~\cite{schmitt2016self,bottcher2020new}. The investigation of QDs has become an intriguing domain, whereby their integrity is preserved through the interaction of quantum fluctuations and mean-field dynamics~\cite{su2024ground,bottcher2020new}.The stability of the droplets are determined via modulation instability (MI) analysis \cite{veni2024numerical,veni2024exploring,tabi2025coupled}. This balance results in the emergence of a new phase of matter, in which the droplet can exist without any external confinement, much like a liquid droplet in free space. Meanwhile, the investigation of superfluidity ~\cite{xi2016droplet,wachtler2016ground} was also examined in addition to droplet formation. 
Vortex states in BECs~\cite{heyl2022vortex,klaus2022observation} are of considerable interest due to their robust nature and the intricate patterns they form. These vortices can induce rich dynamical behaviour and complex spatial structures when embedded in a quantum droplet. However, stabilizing vortex states is a challenging task~\cite{cidrim2018vortices}. Understanding the formation, stability, and interactions of vortex states within (2+1) dimensional quantum droplets provides profound insights into non-linear quantum dynamics, topological phase transitions, and the role of dimensionality in quantum many-body systems. The addition of SOC in BECs has generated significant interest in studying the spatially inhomogeneous states of binary components, as its interaction with the nonlinearity enables the prediction of diverse nonlinear phenomena~\cite{meng2016experimental,zhang2016properties,zhong2018self,tabi2021generation}. Specifically, stable 2D matter-wave solitons sustained by SOC in binary Bose-Einstein condensates, manifested as mixed modes and semi-vortices \cite{Ben-Li}, can be replicated through the spatiotemporal propagation of light in a dual-core nonlinear optical waveguide, where SOC is depicted by the temporal dispersion of the inter-core coupling \cite{Konotop}. Recent  studies have demonstrated that SOC can play a significant role in modifying the structure and stability of quantum droplets in both Bose and Fermi systems. In particular, SOC-induced modifications of droplet phases have been investigated in spinor Bose gases \cite{li2017two} and two-component Fermi systems \cite{cui2018spin}, highlighting the richness of emergent phases in such settings. In parallel with these developments, the significance of SOC effects has been recognized for their role in Exciton-polariton condensates (EPCs)~\cite{sakaguchi2017spin,
aristov2022screening,li2022manipulating,whittaker2018exciton,klaas2019nonresonant}. Exciton-polaritons are hybrid quasiparticles that arise from the strong coupling between excitons (bound electron-hole pairs) and photons confined within microcavities~\cite{jheng2022spontaneous,jiang2022exciton,li2021theory} or photonic structures~\cite{carusotto2013quantum}.
These distinctive entities encapsulate the buoyancy and coherence of photons, along with the robust interactions of excitons, making them an intriguing platform for exploring quantum processes in solid-state systems. Since their initial experimental manifestation, exciton-polaritons have opened up novel avenues for investigating macroscopic quantum coherence~\cite{luders2021quantifying}, non-equilibrium condensates~\cite{pieczarka2020observation}, and quantum fluid dynamics~\cite{peng2022room}, even at elevated temperatures. Exciton-polaritons have been utilized to investigate topological states in photonic lattices. Coupled waveguide lattices have demonstrated resilient topological edge and corner states, facilitating the development of innovative optoelectronic devices~\cite{wen2025plasmon}. Due to the rapid propagation and significant nonlinear interactions, polaritons offer a fertile platform for exploring high-speed switches~\cite{feng2021all}, routers~\cite{marsault2015realization}, logic gates~\cite{mirek2021neuromorphic,sannikov2024room,li2024all}, and polariton lasers~\cite{fraser2016physics}. On the other hand, incoherently pumped polariton condensates have revealed novel mechanisms for generating and cloning dark solitons~\cite{hu2024dark}, offering a deeper understanding of nonequilibrium dynamics. The characteristics of correlation and fermionization in a 1D exciton-polariton gas can be detected using advanced technology, whereby a series of excitonic quantum dots is linked to a photonic waveguide, featuring a low density of polaritons~\cite{knorzer2024fermionization}.  A recent study \cite{ma2020realization} illustrated the effective management of a polariton vortex, utilizing optical imprinting \cite{lagoudakis2008quantized,kwon2019direct,dall2014creation,munoz2020long}, thereafter transitioning to its orthogonal configuration with counter circulation. A method for activating qubit gates using the polariton vortex was subsequently disclosed \cite{ricco2024qubit}. Although polariton vortices have been proposed for topologically safeguarded classical and quantum information processing \cite{kavokin2022polariton,ricco2024qubit}, the prospective uses of vortex–antivortex pairings explored in \cite{wang2024controllable}.
This study explores the intricate physics of quantum droplets and vortex states in (2+1)-dimensional polariton condensates, focusing on the effects of atomic interactions and spin coupling. In lower dimensions, particularly in (2+1) dimensions, the interplay between reduced dimensionality and quantum effects often gives rise to novel phenomena absent in three-dimensional systems. A key aspect of this study is the role of vortex states—topological excitations characterized by quantized angular momentum and rotational symmetry. These vortices significantly influence the dynamics and stability of quantum droplets, leading to complex interplay between quantum fluctuations and emergent collective behaviors.
To investigate these effects, we employ theoretical models and numerical simulations formulated via coupled Gross-Pitaevskii equations (GPE). Specifically, we examine the formation conditions of QDs in the presence of SOC and quantum fluctuations. Our findings contribute to a deeper understanding of low-dimensional quantum fluids and have potential implications for quantum information processing and novel quantum technologies~\cite{PhysRevB.107.184502}. An overview of the paper’s structure is given below: 
Section 2 presents the numerical study of polariton droplets and vortices, analyzing their formation and stability and
Section 3 summarizes our conclusions and discusses potential future directions.
\section{The model}
The dynamics of a two-component polariton condensate, influenced by Photonic SOC and quantum fluctuations, are articulated by the 2D coupled Gross-Pitaevskii equation \cite{madimabe2023modulational}
 \begin{eqnarray}
i\frac{\partial \psi_l}{\partial t} &=& \bigg[-\frac{1}{2} \nabla^2 +\frac{\delta g}{2}(|\psi_l|^2 + |\psi_{3-l}|^2)\nonumber\\
&&+(-1)^l g (|\psi_l|^2 + |\psi_{3-l}|^2)+ (gn_{l}-\bar{g_{l}}n_{3-l})\nonumber\\
&&- \sqrt{\frac{\sigma_l}{\sigma{3-l}}}p \log p\bigg] \psi_l\nonumber+i\bigg(\frac{Rn_{m}}{2}-\gamma _{c}\bigg)\\  
&&- \sigma \left(\frac{\partial}{\partial x} - i(-1)^l\frac{\partial}{\partial y}\right)^2 \psi_{3-l},\;\;l=1,2.
\label{gpe}
\end{eqnarray}
The excitonic reservoir rate equation coupled to the GP equations are given by 
\begin{equation}
\frac{\partial n_{m}}{\partial t}=P_{m}-\left( \gamma _{R}+R|\psi
_{m}|^{2}\right) n_{m},\quad m=1,2.  \label{rate}
\end{equation}
where the wave functions  $\psi_{1}$ and  $\psi_{2}$ are dimensionless, corresponding to spin-up and spin-down components of the polaritons. $\nabla^2$ is the Laplacian operator in two dimensions.
\begin{figure*}
\centering
\xincludegraphics[width=0.4\linewidth,label=\tiny{a)}]{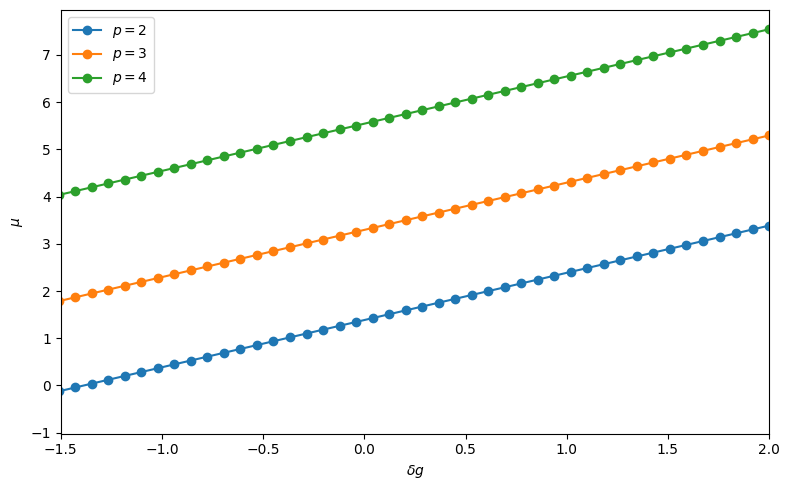}\ %
\xincludegraphics[width=0.4\linewidth,label=\tiny{b)}]{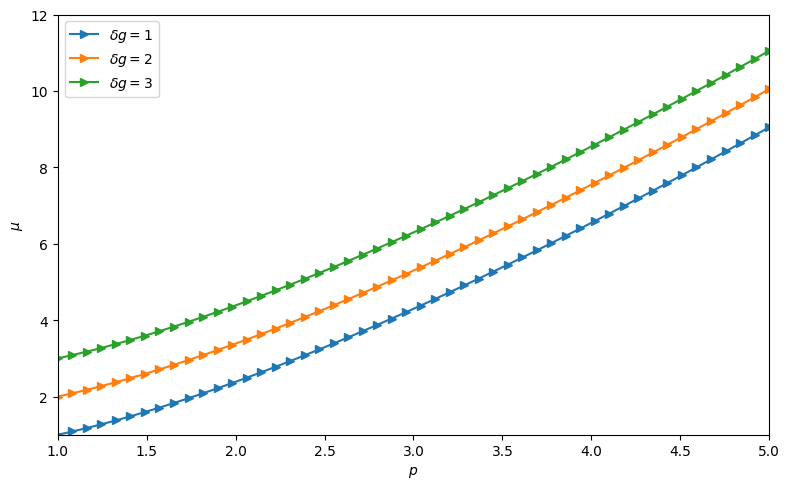} 
\caption{Chemical potential $\mu$ vs $\delta g$ for different values of parameter $p$ are depicted in a). b) Chemical potential $\mu$ vs parameter $p$ for different values of atomic interaction $\delta g$, exhibiting a nonlinear behavior.}
\label{mu_1}
\end{figure*}
The interaction components of the polaritons are defined by $g$ and $\delta g = g + g_{12}$, where $g_{12}$ represents a nonlinear interaction between the intercomponents between exciton-polaritons. The sixth term in Eq. (\ref{gpe}) contributes to the aggregate effects of incoherent pumping, where R indicates the scattering rate of excitons from the reservoir density $n_{m}$ into the condensate, and $\gamma_{c}$ represents the decay rate of polaritons from the condensates, reflecting their lifetime. $P_{m}$ represents the pumping rate, which indicates the external driving force of the system. $\sigma$ denotes the intensity of TE-TM splitting in the polariton modes, commonly referred to as photonic effective spin-orbit interaction. The coupling strength is expressed as $\sigma=\frac{m_{t} - m_{l}}{m_{t}+m_{l}}$ with $m_{t},m_{l}$ are the mass of polaritons in transverse and longitudinal axes. TE-TM splitting in polariton systems results in anisotropic dispersion and spin-dependent coupling, enhancing the system's dynamics and facilitating the investigation of spin-textured patterns and polarization vortices. 
In addition to mean-field contact interactions, we incorporate quantum fluctuation effects via the LHY correction. The expression for the parameter p is given by $p=\frac{1}{2\sqrt{\sigma_1 \sigma_2}}(\sigma_1 |\psi_1|^2+\sigma_2 |\psi_2|^2)$ representing quantum fluctuations \cite{otajonov2022modulational}. The coefficients $\sigma_1$ and $\sigma_2$ represent the relative strength of the LHY term for components 1 and 2, respectively. 
We establish a critical condition in which the excitonic reservoir density $n_{m}$ attains a steady-state value, whereby the pumping from the reservoir ($R n_{m}$) precisely offsets the polariton decay rate ($\gamma_{c}$) 
\cite{zeng2022families}.
\begin{equation}
\frac{\partial n_m}{\partial t} = 0 \quad \Rightarrow \quad n_m = \frac{P_m}{\gamma_R + R|\psi_m|^2}.
\end{equation}
Under the critical balance condition where the gain matches the loss,
\begin{equation}
\frac{R n_m}{2} = \gamma_c,
\end{equation}
we obtain a constant reservoir density:
\begin{equation}
n_m = \frac{2 \gamma_c}{R}.
\end{equation}
Thus, the gain-loss term $i(Rn_m/2 - \gamma_c)$ vanishes, and the dynamics become conservative.
Furthermore, we concentrate on the regime where the polariton-polariton interaction predominates over the reservoir interaction \cite{sakaguchi2017spin}.
\begin{equation}
g(|\psi_1|^2 + |\psi_2|^2) \gg g_R n_l,\; \bar{g}_R n_{3-l},
\end{equation}
and hence we neglect the terms $g n_l - \bar{g}_l n_{3-l}$.
\newline
\newline
In this limit, the system effectively transitions to a conservative regime where the reservoir density remains constant, and the gain-loss dynamics no longer explicitly affect the evolution. Consequently, Eqs.~(1) and (2) reduce to a closed Gross–Pitaevskii-type model, given by Eq.~(3). This reduction enables us to focus on the intrinsic mechanisms driving vortex and droplet formation due to spin-orbit coupling and nonlinear interactions, while excluding dissipative or pumping-driven effects. Although reservoir dynamics are critical in open polaritonic systems, here we isolate the equilibrium-like behavior of the condensate in the absence of gain and loss. This approach aligns with earlier works modeling conservative condensate behavior in polariton systems \cite{sakaguchi2017spin,zeng2022families}, and allows for a clearer interpretation of self-bound droplet states.
\begin{eqnarray}
i\frac{\partial \psi_{l}}{\partial t}&=&\bigg[-\frac{1}{2}\nabla^{2}+\frac{\delta g}{2}\left(|\psi_{l}|^{2}+|\psi_{3-l}|^{2}\right)\nonumber \\
&&+(-1)^{l}g\left(|\psi_{l}|^{2}-|\psi_{3-l}|^{2}\right)
\\ &&- \sqrt{\frac{\sigma_l}{\sigma{3-l}}}p \log p\bigg]\psi_{l}
+\sigma\bigg(\frac{\partial}{\partial x} \nonumber\\&&+i(-1)^{l}\frac{\partial}{\partial y}\bigg)^{2}\psi_{3-l}\nonumber.
\label{egpe}
\end{eqnarray}

This equilibrium state facilitates the stabilization of condensate density, leading to energy-conserved dynamics. The dimensionless form of the conservative Gross-Pitaevskii equation is represented by Eq. (3) and is scaled by length normalized to $l=\sqrt{\frac{\hbar}{m\omega_{\perp}}}$, energy measured in $\hbar \omega_{\perp}$, time in units of $\omega^{-1}$, and density ($|\psi|^{2}$) in terms of $l^{-1}$.
The normalized number of particles (norm) in the droplets is given by 
\begin{eqnarray}
N=\int_{-\infty}^{+ \infty}\left(|\psi_{1}|^{2}+|\psi_{2}|^{2} \right)dx dy =N_{1}+N_{2},
\end{eqnarray}
where $N_{1}$ and $N_{2}$ are the first and second components' norms, indicating that the total system remains conserved.
Quantum droplets in our model emerge from the balance between mean-field attraction and repulsive quantum fluctuations, as captured by Eq.~(3). However, the exact mechanism by which this balance is achieved can vary depending on the dominant parameters. In some regimes, droplets are stabilized when the intra-component interaction \(g\) is large and the inter-component term \(\delta g\) remains relatively small, leading to localized self-bound states primarily shaped by intra-species nonlinearities. In contrast, other cases show that when \(g\) is small but \(\delta g\) is large, strong inter-component interactions dominate the dynamics and support droplet formation through inter-species coupling. Additionally, when the spin-orbit coupling strength \(\sigma\) is large, it contributes significantly to the stabilization and structuring of the condensate, inducing vortex-like textures even when the nonlinear interactions alone may not support a droplet phase. These varying regimes highlight that droplet formation in this system does not follow a single universal condition but can be realized through different interaction pathways, as confirmed by our simulations. This diversity in balance mechanisms allows for rich dynamical behavior and tunability across the droplet and vortex-droplet phases.
We consider a plane wave solution for the Gross-Pitaevskii equation Eq. (3), such that
\begin{equation}
\psi_{l} = A_{l}e^{-i\mu t}
\end{equation}
The chemical potential can be expressed in the form 
\begin{eqnarray}
\mu &=&\frac{\delta g}{2} \big(A_{l} ^2 + A_{3-l}^2\big) - (-1)^l g \big(A_{l}^2 - A_{3-l}^2 \big)\nonumber\\
&&+ \sqrt{\frac{\sigma_{l}}{\sigma_{3-l}}} p \ln(p)    
\end{eqnarray}

\section{Numerical Results}
Using the split-step Fourier method ~\cite{abdolabadi2023split} the Gross-Pitaevskii equation was solved with real-time propagation. The numerical resolution was adjusted based on the interaction parameters: for weak interactions ($g$,$\delta g$), larger time steps and fewer Fourier modes were used due to slower system evolution. For strong interactions ($g$,$\delta g$), finer time steps and higher Fourier mode numbers were required to accurately resolve rapid density variations and strong nonlinear effects.
The interaction strengths g and $\delta g$ in our simulations are given in dimensionless form, normalized by the characteristic harmonic oscillator length and trap frequency of the system. This normalization can result in large numerical parameter values while maintaining correspondence to experimentally accessible physical regimes when appropriately rescaled. Experimentally, strong effective nonlinearities of this magnitude can be realized by tuning scattering lengths near Feshbach resonances or by increasing the number of atoms loaded into the trap.

The Gaussian initial state used to study the dynamics are given by 
\begin{eqnarray}
    \psi_{1,2} = e^{-(x^2+y^2)/\alpha^{2}}
\end{eqnarray}
where $\alpha$ is the width parameter. The presented numerical results correspond to early-time dynamical states, obtained by direct real-time simulations of the reduced conservative Gross–Pitaevskii equation [Eq.~(3)]. The initial conditions consist of real-valued Gaussian profiles as given in Eq.~(7), with no imposed phase winding. The vortex and droplet structures shown in the figures emerge dynamically due to nonlinear interactions and spin-orbit coupling. These are not stationary solutions, but transient, self-organized configurations resulting from the interplay of interaction and SOC terms. Meanwhile, the dynamics associated with stationary states are explored in \cite{sanjay2025vortex}.
The coupling of photonic spin-orbit interactions with intra- and inter-component interactions generates vortex. These vortex-like formations merge to form a large vortex, characterized by increasing values of photonic spin-orbit coupling. Vortices can be identified on phase diagrams by detecting phase singularities at points where the real-space density is zero, indicating the presence of a vortex core. The phase encircling the singularities intensifies with stronger inter-component interactions, although low inter-component interaction levels exhibit independent behavior between the two components. At higher intra-component contact strengths, this interaction leads to the collapse of the condensate and the formation of QDs.
\begin{figure*}
\centering
\xincludegraphics[width=0.8\linewidth,label=\tiny{a)}]{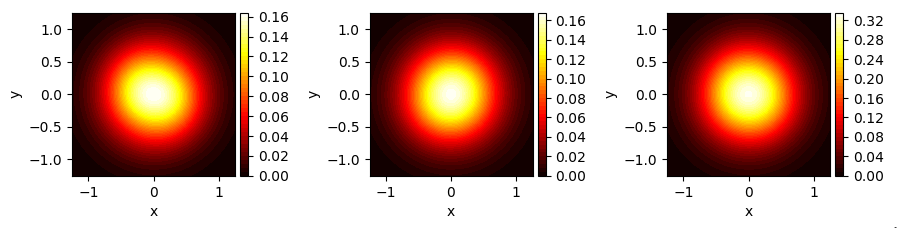}\ %
\xincludegraphics[width=0.8\linewidth,label=\tiny{b)}]{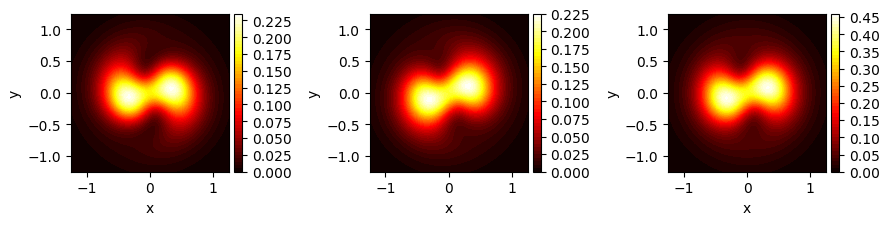}\ %
\xincludegraphics[width=0.8\linewidth,label=\tiny{c)}]{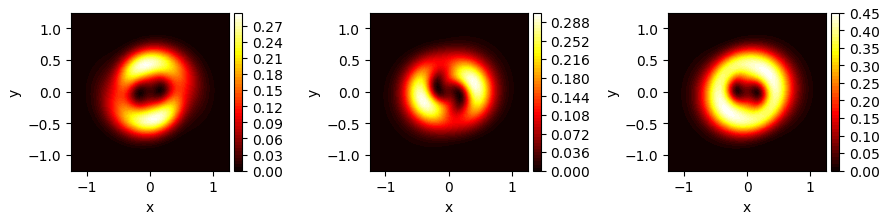}\ %
\caption{Visible vortices are observed for increasing interaction parameters. Contour plot densities $|\psi_{1}(x,y)|^{2},|\psi_{2}(x,y)|^{2}$ and $|\psi_{1}(x,y)|^2+|\psi_{2}(x,y)|^2$ subjected for different interaction coefficients, illustrated in a) for $\delta g=500$, $g=100$ ; b) for $\delta g=5500$, $g=100$ and c) $\delta g=17500$, $g=500$. The spin strength ($\sigma$) illustrating the density profiles are given by $\sigma=0.5$.}
\label{fig:1}
\end{figure*}
\begin{figure*}
\centering
\xincludegraphics[width=0.8\linewidth,label=\tiny{a)}]{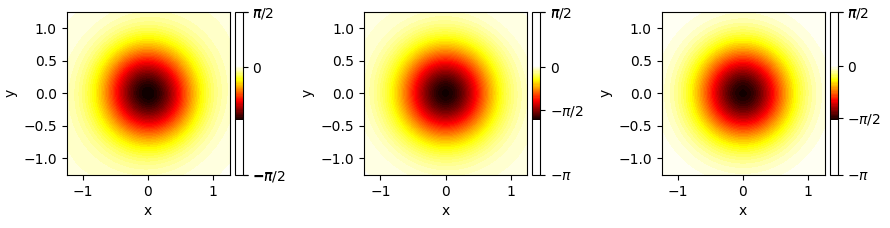}
\xincludegraphics[width=0.8\linewidth,label=\tiny{b)}]{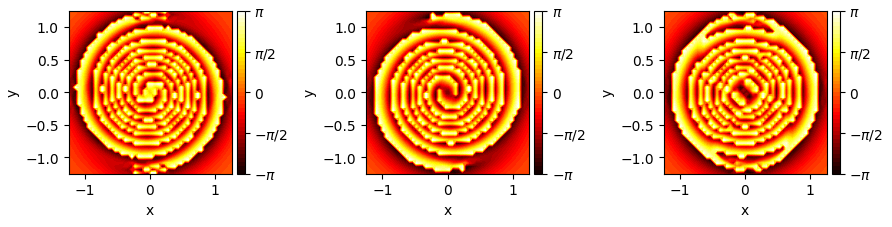}
\xincludegraphics[width=0.8\linewidth,label=\tiny{c)}]{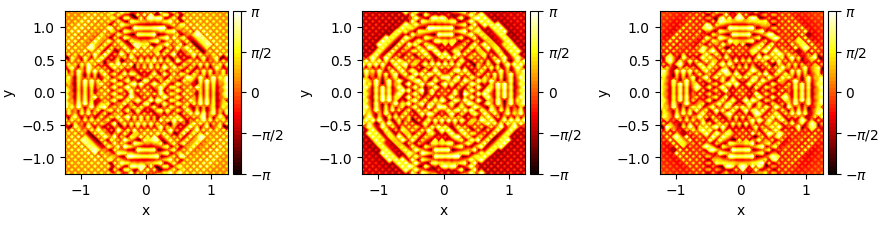}
\caption{Phase profiles showing the emergence of singularities as the intercomponent interaction $\delta g$ increases, with the intracomponent interaction and spin strength fixed at $g = 500$ and $\sigma=0.5$.
(a) For $\delta g = 1000$, a smooth and nearly uniform phase profile is observed, with no visible singularities.
(b) As $\delta g$ increases to $17500$, spiral phase windings appear, indicating the formation of phase singularities associated with droplet interactions.
(c) For $\delta g = 57500$, the spiral windings become complex and the fringes start to dissolve.
}
\label{fig:phase}
\end{figure*}
\begin{figure*}
\centering
\xincludegraphics[width=0.8\linewidth,label=\tiny{a)}]{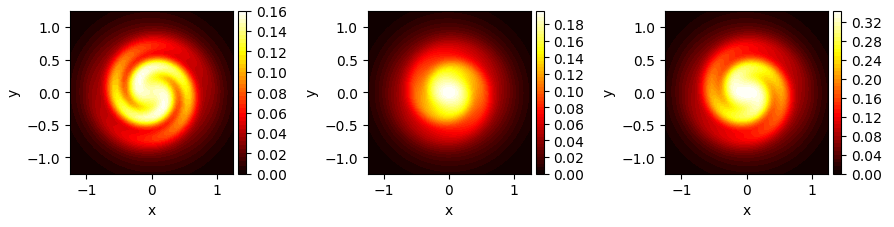}\ %
\xincludegraphics[width=0.8\linewidth,label=\tiny{b)}]{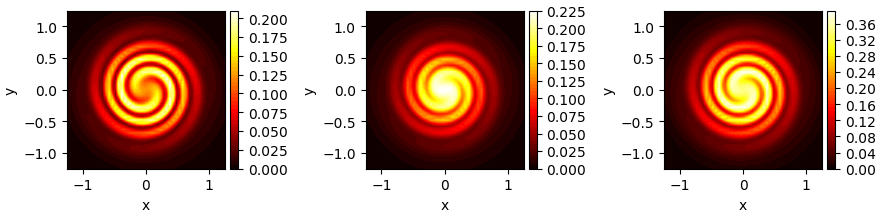}\ %
\xincludegraphics[width=0.8\linewidth,label=\tiny{c)}]{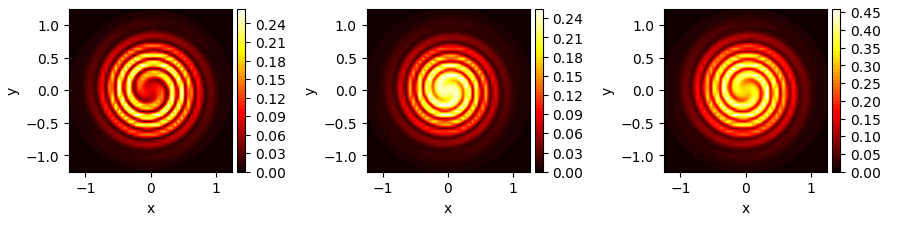}\ %
\caption{Vortex pairs shown for fixed spin interaction  $\sigma=0.5$ and different atomic interaction $g$ while $\delta g$ remaining constant, illustrated through the density distribution  viz., $g = 2000$ (a), $g = 4000$ (b), and $g = 6000$ (c) with $\delta g=500$ in all the cases.}
\label{fig:2}
\end{figure*}
The emergence of quantum droplets in our simulations is consistent with earlier works on self-bound states in dipolar and two-component condensates \cite{petrov2015quantum,chomaz2016quantum}. These studies have shown that quantum fluctuations stabilize these droplets by counteracting attractive interactions, a mechanism also observed in our numerical findings. However, unlike prior studies that focused on dipolar BECs, our work emphasizes the interplay between spin-orbit coupling and interaction parameters ($\sigma$, $\delta g$ and g), revealing a tunable transition between vortex and droplet phases. This highlights a novel mechanism for controlling self-bound quantum states, which has not been extensively explored in previous literature.

We initially investigate the action of chemical potential $\mu$ vs atomic interaction for various values of $p$, shown in Fig. \ref{mu_1}a, where $\mu$ exhibits linear behavior even for different values of $p$. In Fig. \ref{mu_1}b, the effect on chemical potential $\mu$ by $p$ for different values of $\delta g$ is observed. In this case, $\mu$ increases as the parameter $p$ increases. The dynamics of vortices for a fixed value of spin strength $(\sigma)$, while varying the interaction term $g$ and $\delta g$ exhibits a progressive evolution from stability (no vortex) to the formation of a vortex-like structure shown in Fig. \ref{fig:1}. 
\begin{figure*}
\centering
\xincludegraphics[width=0.8\linewidth,label=\tiny{a)}]{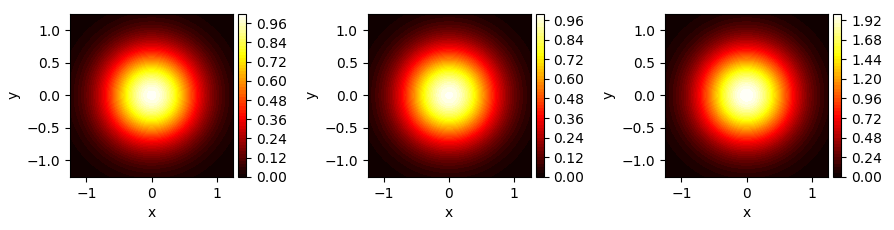}\ %
\xincludegraphics[width=0.8\linewidth,label=\tiny{b)}]{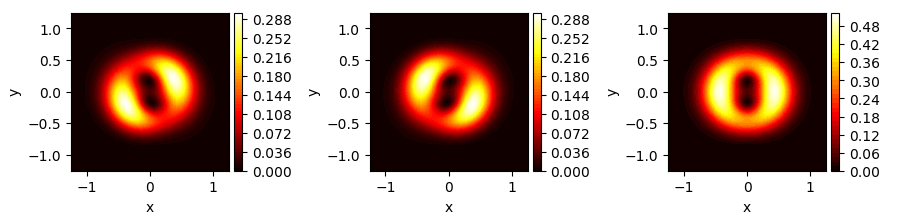}\ %
\xincludegraphics[width=0.8\linewidth,label=\tiny{c)}]{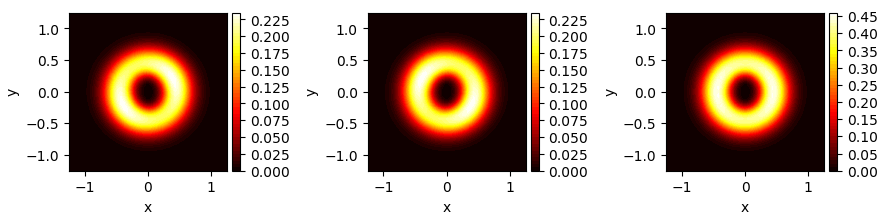}\ %
\caption{Formation of droplet and vortex state for fixed atomic interaction $g=100$ and $\delta g=500$ and different spin strength $\sigma$. A droplet state is observed for $\sigma=0.05$ in (a). The transition from droplet state observed as spin strength increases: 
$\sigma=2$ in (b) and $\sigma=11$ in (c).}
\label{fig:3}
\end{figure*}
For interaction parameters, $\delta g = 500$, $g = 100$ and spin strength, $\sigma = 0.5$ with fluctuation parameters $\sigma_{1} = 1$, $\sigma_{2} = 1$, the BEC undergoes crystallization, resulting in a stable density profile devoid of discernible vortices or singularities, as depicted in Fig. \ref{fig:1}a). The density distributions $(|\psi_{1}|^{2}, |\psi_{2}|^{2})$ remain uniform, suggesting a near-ground-state configuration with minimal influence from spin-orbit coupling. The density distributions $(|\psi_{1}|^{2}, |\psi_{2}|^{2})$ remain uniform, suggesting a near-ground-state configuration with minimal influence from spin-orbit coupling. Increasing the interactions to $\delta g = 5500$ and $g = 100$ leads to the onset of droplet splitting accompanied by rotational motion, observed in Fig. \ref{fig:1}b). At even higher interaction strengths, specifically $\delta g = 17500$ and $g = 500$ in Fig. \ref{fig:1}c), rapid rotational dynamics emerge, culminating in the formation of vortex-like structures. Upon tuning the interaction parameters $\delta g$ and $g$, the system undergoes spontaneous vortex nucleation, with the vortex structures becoming more pronounced at stronger nonlinearities. These findings underscore the crucial role of interaction-induced instabilities in driving the condensate into topologically nontrivial states characterized by vortex lattices and density modulations.

The behavior of phase profiles under the action of increased atomic interaction are shown in Fig. \ref{fig:phase}). In Fig. \ref{fig:phase}a), corresponding to $\delta g = 1000$, the phase remains smooth and nearly uniform across the system, indicating the absence of phase singularities or topological defects. As $\delta g$ increases to $17500$ Fig. \ref{fig:phase}b), spiral phase windings emerge, signaling the formation of phase singularities and the onset of inter-droplet phase coupling, confirming the emergence of vortex behavior in this regime. Increasing the interaction strength to $\delta g = 57500$ Fig. \ref{fig:phase}c) leads to more complex structures and fringes are begins to dissolve.
\begin{figure*}
\centering
\xincludegraphics[width=0.8\linewidth,label=\tiny{a)}]{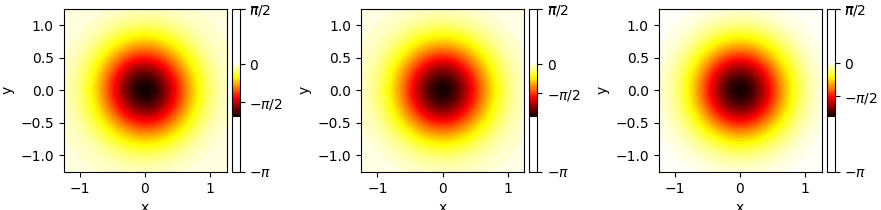}\ %
\xincludegraphics[width=0.8\linewidth,label=\tiny{b)}]{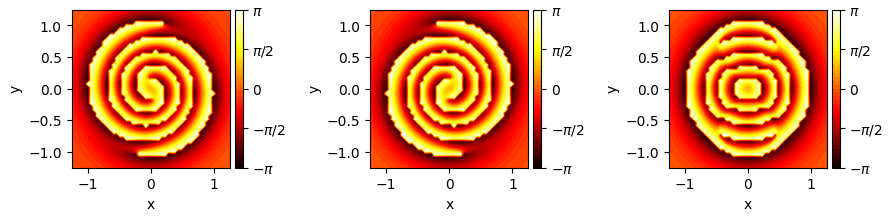}\ %
\xincludegraphics[width=0.8\linewidth,label=\tiny{c)}]{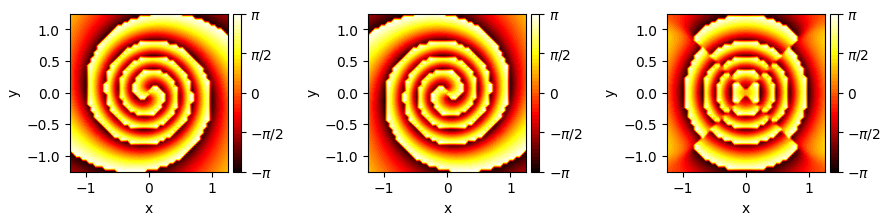}\ %
\caption{Phase profiles indicating the emergence of singularities as the spin interaction $\sigma $ increases, with the atomic interaction fixed at $\delta g = 500$ and $g=100$.
(a) For $\sigma = 0.5$, a smooth and dense phase profile is observed.
(b) As $\sigma$ increases to $2$, spiral phase windings appear.
(c) For $\sigma=100$, the enhanced spiral windings are observed representing the vortex behavior.} 
\label{phase_spin}
\end{figure*}
The manifestation of vortex droplets under varying atomic interaction strengths, with fixed coupling parameters, is demonstrated in Fig. \ref{fig:2}. The interplay between the internal spin degrees of freedom and the atom cloud is influenced by the coupling parameter. The vortex pair configurations are examined for a fixed spin interaction parameter $\sigma = 0.5$, while the atomic interaction strength $g$ is varied, with the intercomponent interaction $\delta g$ held constant ($\delta g = 500$). The density distributions $|\psi|^{2}$ are shown for three representative values of $g$: a) $g=2000$, b) $g=4000$, and c) $g=6000$. Each panel corresponds to a different combination of interaction strengths, illustrating the evolution of vortex structures as $g$ increases.
 In all panels, the vortices are identified by their characteristic low-density spiral cores, surrounded by density oscillations that reflect underlying quantum interference and nonlinear effects. The systematic increase in $g$ leads to the enhancement of nonlinear interactions, resulting in sharper and more pronounced vortex spirals in the density distribution. By keeping $\delta g$ fixed, the influence of $g$ is isolated, providing a clearer understanding of how intra-atomic scattering affects the topological excitations of the condensate. It is also noteworthy that, despite maintaining a constant spin interaction strength, the increase in atomic interaction parameters results in an enhanced spiral structure of the vortex. This evolution in the vortex's spiral pattern can be attributed to the strengthening of nonlinear interactions, which modulate the topological structure and dynamics of the vortex core. The increased interaction strength leads to a sharper and more defined spiral, reflecting the influence of intra-atomic scattering on the condensate's topological excitations.

Now the investigation of droplet and vortex states for a fixed atomic interaction strength $g = 100$, $\delta g = 500$ and different spin interaction parameters $\sigma$ are illustrated in Fig. \ref{fig:3}. For a low spin interaction strength ($\sigma = 0.05$) shown in Fig. \ref{fig:3}a), the system stabilizes into a droplet state characterized by a high-density central peak and the absence of phase singularities. This configuration corresponds to a self-bound quantum droplet, stabilized by beyond-mean-field effects, without topological excitations. As the spin interaction is increased to $\sigma = 2$ in Fig. \ref{fig:3}b), the droplet state transitions to a vortex configuration, exhibiting a density depletion (vortex core) at the center. This transition marks the emergence of topological excitations due to spin-dependent nonlinearities, leading to phase winding and angular momentum quantization in the condensate. For a further increase in $\sigma$ to 11, well-defined vortex rings are observed in Fig. \ref{fig:3}c), with density minima. These features are indicative of stable vortex states that arise in the presence of strong spin interactions. The interaction between SOC and atomic forces leads to this transition, where $\sigma$ is the pivotal factor. Each vortex represents a topological defect associated with a specific winding number. The progression from droplet to vortex state highlights the critical influence of spin interactions in shaping the condensate’s topology and stability.

\begin{figure*}
\centering
\xincludegraphics[width=0.48\linewidth,label=\tiny{a)}]{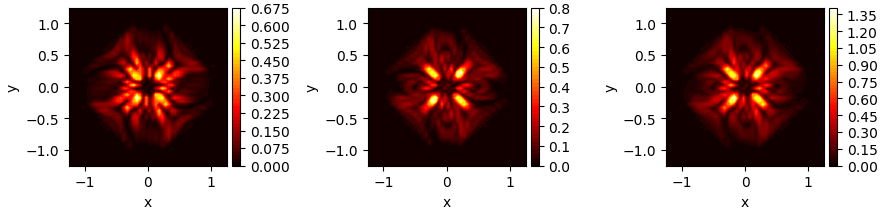}$\quad$
\xincludegraphics[width=0.48\linewidth,label=\tiny{}]{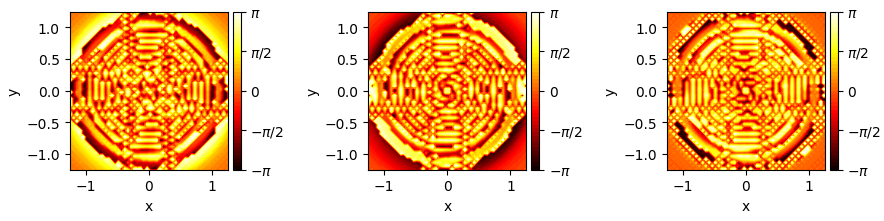}\ %
\xincludegraphics[width=0.48\linewidth,label=\tiny{b)}]{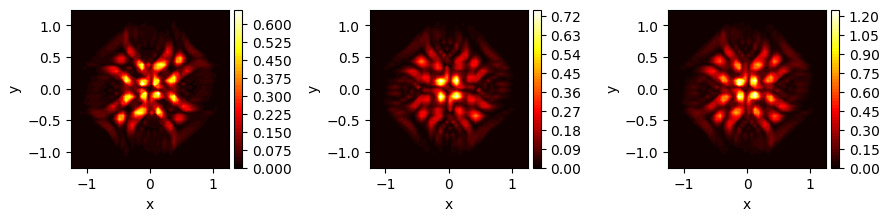}$\quad$
\xincludegraphics[width=0.48\linewidth,label=\tiny{}]{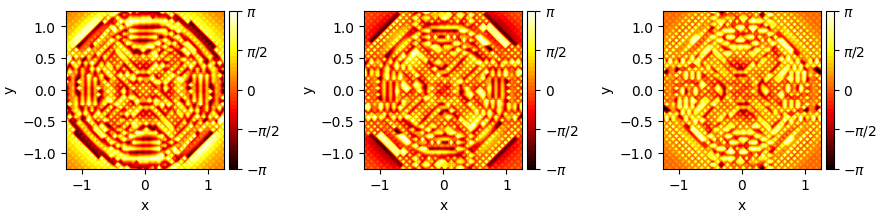}\ %
\xincludegraphics[width=0.48\linewidth,label=\tiny{c)}]{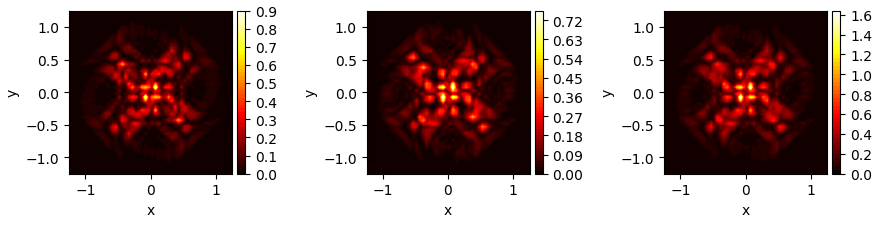}$\quad$
\xincludegraphics[width=0.48\linewidth,label=\tiny{}]{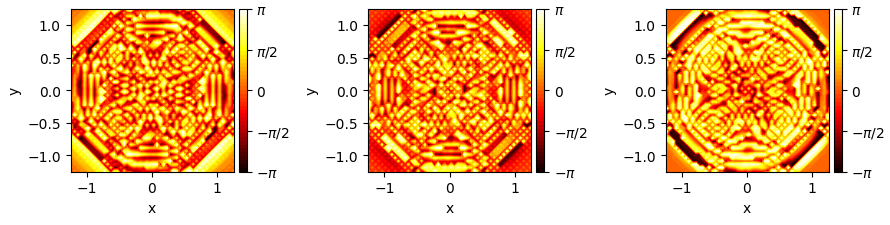}\ %
\caption{Quantum droplets arranged in a lattice state for fixed spin interaction $\sigma=200$ and intra-component interaction 
$\delta g=6500$. The effect of increasing inter-component interaction is illustrated: 
$g=20000$ in (a), 
$g=30000$ in (b), and 
$g=50000$ in (c). The phase plots in the right panel highlight intricate phase textures and singularities associated with droplet interactions. As $g$ increases, the system exhibits a stable droplet configuration with reduced coherence, as evidenced by phase noise in the high-interaction regime.}
\label{fig:4}
\end{figure*}

\begin{figure*}
\centering
\xincludegraphics[width=0.49\linewidth,label=\tiny{a)}]{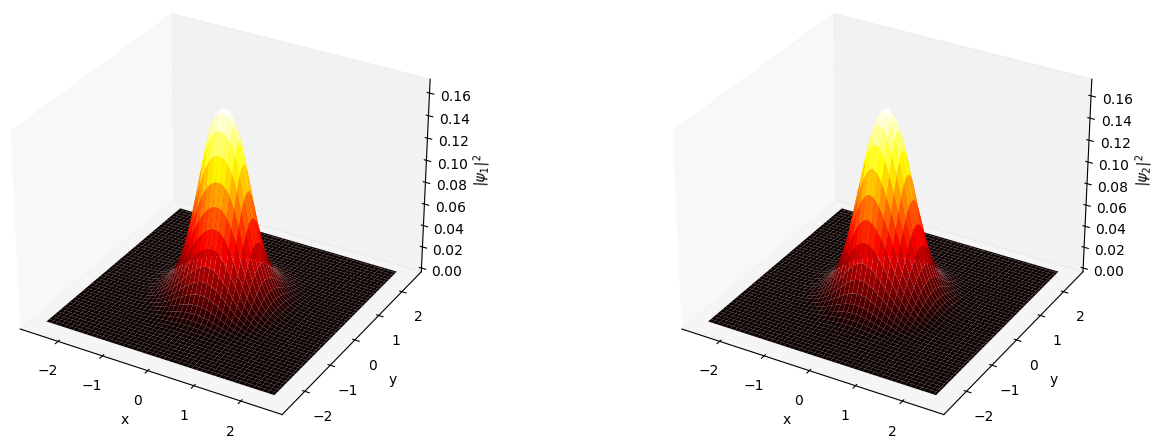}
\xincludegraphics[width=0.49\linewidth,label=\tiny{b)}]{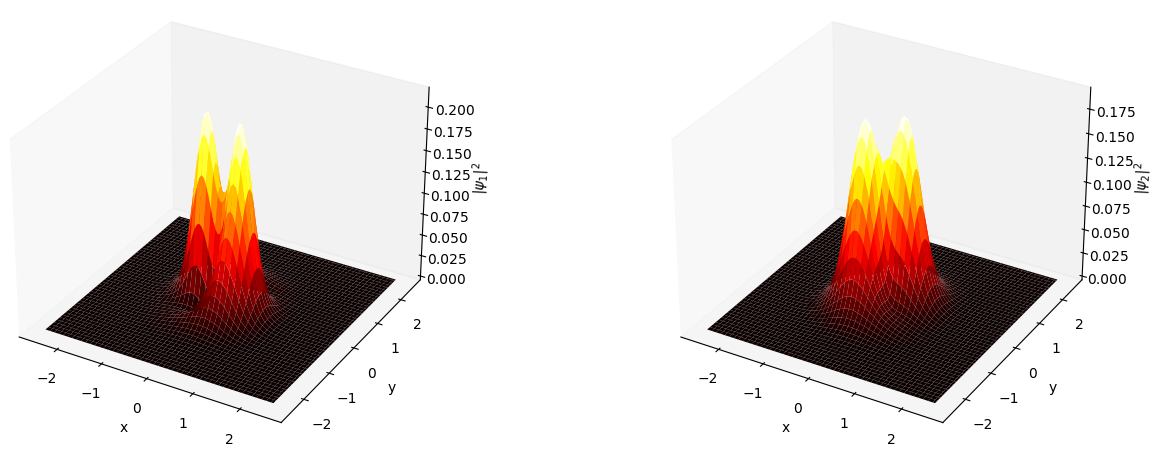}
\caption{Three-dimensional density profiles of a quantum droplet under interaction parameters $\delta g=500$ and $g=100$. (a) shows stable, symmetric density peaks characteristic of a well-formed droplet for $\sigma-1.5$, while (b) illustrates variations in droplet shape and peak due to changes in interaction strength, highlighting the sensitivity of droplet structure to the parameter $\sigma=15$.}
\label{fig:5}
\end{figure*}

As the spin interaction \(\sigma\) increases, the phase profile of the system shown in Fig. \ref{phase_spin} undergoes significant changes, revealing the emergence of singularities in the wavefunction. For fixed atomic interaction strengths \(\delta g = 500\), \(g = 100\) and $\sigma=0.05$, the system exhibits a smooth phase profile. This behavior suggests that the spin-orbit coupling is weak, with no discernible singularities or vortex-like structures present in the phase, as can be seen in Fig. \ref{fig:3}a).  For \(\sigma = 9\), as the spin interaction increases,  spiral phase windings begin to form.
For \(\sigma = 100\), a significant enhancement in the spiral phase windings is observed. This corresponds to the further development of vortex structures, where the phase profile clearly reflects the formation of vortices. This behavior highlights the strong influence of spin-orbit coupling on the system's phase structure. Therefore, as \(\sigma\) increases, the system exhibits a transition from a smooth, to spiral phase, characterized by distinct vortex singularities, underscoring the pivotal role of spin interactions in shaping the system's topological properties.
Further, we demonstrate the formation of quantum droplets arranged in a lattice configuration under various interaction conditions, with particular emphasis on how the inter-component interaction strength $g$ influences the system. In this study, we fix the following parameters:
Spin interaction parameter $\sigma = 200$
and $\delta g = 6500$. As the atomic  interaction $g$ increases, we observe a clear evolution in the density distribution and phase textures of the quantum droplets. 
At $g = 20000$, Fig. \ref{fig:4}a), the density distribution reveals a lattice of quantum droplets, characterized by intricate and complex patterns. The phase distribution associated with these droplets shows significant complexity, with numerous singularities indicating substantial quantum coherence fluctuations and correlations in the system. Upon increasing the inter-component interaction to $g = 30000$ (Fig. \ref{fig:4}b), the density pattern of the droplets becomes more defined and stable. The phase textures show a noticeable reduction in singularities, reflecting a suppression of quantum noise and phase disorder, and indicating an improvement in the system's coherence and stability. For a further increase to $g = 50000$ (Fig. \ref{fig:4}c), the droplets density distribution becomes even more regular and stable, with well-formed, well-ordered configurations. The phase profiles exhibit much less noise, with significantly fewer phase singularities, signifying a stable configuration with minimal coherence degradation. The phase textures in the right panels of each density profiles, provide a visualization of the phase distribution across space. The presence of phase singularities (discontinuities in phase) is indicative of quantum noise or a loss of coherence. As $g$ increases, the reduction in these phase singularities suggests that the system evolves toward a more coherent and ordered state, with a higher degree of stability.
Thus, we conclude that controlling the inter-component interaction strength $g$ plays a key role in tuning the coherence and stability of quantum droplets arranged in a lattice. Fig. \ref{fig:5} presents the three-dimensional density profiles of the droplets for \( g\delta g = 500 \) and \( g = 100 \). In Fig. \ref{fig:5}a), corresponding to \( \sigma = 1.5 \), both the spin-up and spin-down components, a single, smooth density peak, indicative of a stable configuration. When \( \sigma \) is increased to 15, the density profile evolves to display two distinct and stable peaks, signaling a transition to a different stable state, observed in Fig. \ref{fig:5}b).
\section{Conclusion}
In this study, we investigated the interplay between photonic spin-orbit coupling and intra- and inter-component interactions in a binary Bose-Einstein condensate (BEC). Our findings reveal a diverse array of emergent quantum phases, ranging from stable condensate states to vortex-dominated and droplet-dominated regimes, governed by the interaction parameters and spin-orbit coupling strength. This work contributes to a broader understanding of topological excitations, quantum phase transitions, and beyond-mean-field effects in spin-orbit-coupled condensates.
At weak spin-orbit coupling ($\sigma = 0.05$), the system remains stable and symmetric, with no vortices or phase singularities in the density profile. As the coupling strength increases, the system undergoes a topological transition, where angular momentum is injected into the condensate, leading to the formation of quantized vortices. These vortices, characterized by localized phase singularities, are clear indicators of macroscopic rotational coherence. For intermediate spin-orbit coupling strengths, the droplet starts to split and spin, forming a vortex-like structure. At sufficiently high values ($\sigma = 11$), a vortex-antivortex pair with a high topological charge, indicating a single topological excitation is observed. The winding numbers and associated phase singularities confirm the system's transition to a coherent macroscopic rotational state.
The competition between spin-orbit coupling and atomic interactions ($\delta g$ and $g$) introduces additional complexity to the system’s phase behavior. At low values $\delta g$ and $g$ where $\delta g \gg g$, the two spin components evolve nearly independently, maintaining a near-circular density distribution. As $\delta g$ increases, inter-component coupling strengthens, leading to symmetry breaking and the formation of vortex-like structure. The presence of vortex is also confirmed by their phase profiles, where an increase in $\delta g$ results in spiral phase windings. Further, an increase in $\delta g$ deforms the vortex, confirmed by the phase structure, which shows dissolved fringes. Similarly, for the case of an increase in the value $g$ while maintaining $\delta g$ and $\sigma$ as constant, causes the condensate to spin. In this regime, vortex cores develop spiral-like arrangements, forming structured patterns that arise due to the balance between repulsive inter-component interactions and attractive intra-component forces. This transition marks a nontrivial interplay between interaction-induced stability and spin-orbit coupling-induced rotation, leading to the emergence of distinct topological configurations within the condensate.
For higher intra-component interaction strengths ($g$), beyond-mean-field effects play a crucial role in stabilizing the condensate. Unlike the vortex-dominated phase, where angular momentum is the defining feature, the droplet phase is governed by competition between quantum fluctuations and attractive interactions. In this regime, density distributions exhibit localized peaks, with droplets forming self-organized lattice structures. Phase portraits indicate a loss of long-range coherence, replaced by localized quantum fluctuations that stabilize the droplets. The emergence of these quantum droplets suggests a transition from a coherent superfluid phase to a self-trapped state, where mean-field interactions dominate kinetic energy contributions.
The presence of significant phase noise in the droplet regime is a signature of the system's transition from a long-range ordered condensate to a localized state, where interactions drive the formation of stable-density droplets. This finding is particularly relevant for understanding the stability and tunability of self-bound states in quantum fluids, which may have implications for designing novel quantum materials and dipolar condensates.

Our study highlights the tunability of quantum vortex and droplet states through external parameters, such as spin-orbit coupling ($\sigma$) and atomic interactions ($\delta g$ and $g$). The ability to control these parameters suggests potential applications in quantum simulations of topological defects and exotic superfluid states, precision sensing using vortex-based macroscopic quantum states, and controlled manipulation of quantum phases for future quantum technologies. Furthermore, the observed phase transitions provide a theoretical platform for exploring fundamental topological and quantum phase phenomena. The ability to drive the condensate from a droplet-dominated to a vortex-dominated regime offers insights into the mechanisms underlying superfluidity, angular momentum quantization, and self-organization in quantum fluids.
Promising directions for future work include exploring the stability of dipolar vortices and quantum droplets, investigating the effects of confining potentials on vortex lattice formation and droplet stabilization, and extending the analysis to (3+1)-dimensional condensates to examine the role of higher-order interactions in quantum phase transitions. Furthermore, incorporating reservoir dynamics and non-conservative effects—such as gain-loss imbalance, incoherent pumping, and reservoir-mediated instabilities—could provide valuable insights into the behavior of vortex quantum droplets in realistic driven-dissipative systems. These aspects, however, are beyond the scope of the present study but offer promising directions for future investigation.


\section*{Acknowledgements}
S. Sanjay and S. Saravana Veni acknowledge Amrita Vishwa Vidyapeetham,
Coimbatore, where this work was supported under Amrita Seed Grant (File
Number: ASG2022141). CBT thanks the Kavli Institute for Theoretical Physics (KITP), University of California Santa Barbara (USA), where this work was supported in part by the National Science Foundation Grant No. NSF PHY-2309135.
\end{document}